\newcommand{\beq}{\begin{equation}}             
\newcommand{\eeq}{\end{equation}}               
\newcommand{\bqry}{\begin{eqnarray}}            
\newcommand{\eqry}{\end{eqnarray}}              
\newcommand{\bqryn}{\begin{eqnarray*}}          
\newcommand{\eqryn}{\end{eqnarray*}}            
\newcommand{\PD}[2]                             
    {\frac{\partial^{#2}}{\partial #1^{#2}}}    

\documentclass[preprint,prl,aps,showpacs]{revtex4}
\usepackage{graphicx}
\usepackage{xcolor}

\begin{document} 



\title{Experimental and theoretical confirmation of an orthorhombic 
phase transition in niobium at high pressure and temperature}

\author{Daniel Errandonea$^1$, Leonid Burakovsky$^2$, Dean~L.~Preston$^3$, 
Simon~G.~MacLeod$^{4,5}$, David~Santamar\'{i}a-Perez$^1$, Shaoping~Chen$^2$, 
Hyunchae~Cynn$^6$, Sergey~I.~Simak$^7$, Malcolm~I.~McMahon$^4$, 
John~E.~Proctor$^4$, and Mohamed Mezouar$^8$}

\thanks{Requests for experimental materials should be addressed to D.E. 
(e-mail: daniel.errandonea@uv.es). Requests for theoretical materials 
should be addressed to L.B. (e-mail: burakov@lanl.gov).}

\affiliation{$^1$ Departamento de F\'{i}sica Aplicada-ICMUV, MALTA 
Consolider Team, Universidad de Valencia, Edificio de Investigaci\'{o}n, 
C/Dr. Moliner 50, Burjassot, 46100 Valencia, Spain \\
$^2$ Theoretical and $^3$ Computational Physics Divisions,
Los Alamos National Laboratory, Los Alamos, New Mexico 87545, USA \\
$^4$ SUPA, School of Physics and Astronomy and Centre for Science at Extreme 
Conditions, The University of Edinburgh, Edinburgh, EH9 3FD, UK \\
$^5$ AWE, Aldermaston, Reading, RG7 4PR, UK \\
$^6$ Physics Division, Lawrence Livermore National Laboratory, Livermore, 
California 94551, USA \\
$^7$ Department of Physics, Chemistry and Biology, Link\"{o}ping University, 
58183 Link\"{o}ping, Sweden \\
$^8$ ID27 Beamline, European Synchrotron Radiation Facility, 71 Avenue des 
Martyrs, 38000 Grenoble, France}





\begin{abstract}
Compared to other body-centered cubic (bcc) transition metals Nb has been 
the subject of fewer compression studies and there are still aspects of its 
phase diagram which are unclear. Here, we report a combined theoretical and 
experimental study of Nb under high pressure and temperature. We present 
the results of static laser-heated diamond anvil cell experiments up to 120 
GPa using synchrotron-based fast x-ray diffraction combined with ab initio 
quantum molecular dynamics simulations. The melting curve of Nb is determined, 
and evidence for a solid-solid phase transformation in Nb with increasing 
temperature is found. The high-temperature phase of Nb is orthorhombic Pnma. 
The bcc-Pnma transition is clearly seen in the experimental data on the Nb 
principal Hugoniot. The bcc-Pnma coexistence observed in our experiments is 
explained. Agreement between the measured and calculated melting curves is 
very good except at 40-60 GPa where three experimental points lie below the 
theoretical melting curve by 250 K (or 7\%); a possible explanation is given.
\end{abstract}

\maketitle

\section*{Introduction}

The d-block transition metals are extremely important technological materials. 
However, their phase diagrams have remained virtually uknown over decades of 
research. More recently, advances in theoretical \cite{8,9,4,3,10,6,2,5,31,Re} 
and experimetal \cite{Ni,KZ,14,13,27,16,39,11,Re-Boehler,15,26,12,Mo,Pt,V} 
techniques started offering the possibility to gain better understanding of 
the high-pressure (HP) and high-temperature (HT) behavior of transition metals 
and to elucidate the fundamental mechanisms responsible for the physical 
appearance of their HP-HT phase diagrams. Of special interest is HP-HT 
polymorphism in the bcc transition metals of Groups 4B (Cr, Mo, W) and 5B 
(V, Nb, Ta) which is related to laser-heated diamond anvil cell (DAC) 
melting experiments bringing up a small slope $(dT/dP)$ of the corresponding 
melting curves in the pressure-temperature $(P$-$T)$ coordinates in 
the megabar $P$ range \cite{Ni,KZ,14,13,27,16,39,Re-Boehler}. These flat 
melting curves contradict the results of both more recent experiments 
\cite{11,15,26,12,Mo,Pt,V} and calculations \cite{9,3,6,5,31,Re}. This 
apparent controversy has caused several hypotheses to be proposed. In 
particular, the flat melting curves could in fact correspond to solid-solid 
phase transition boundaries at HP-HT below melting. Such HP-HT solid-solid 
phase transitions have been predicted for Ta \cite{4,10,31}, Mo \cite{8} and 
Re \cite{Re}. In contrast to other metals, only few studies on Nb have been 
reported in the literature. The initial ($P=0$) slope of the Nb melting curve 
is known from isobaric expansion measurements \cite{20}, although the melting 
curve itself has not been measured. Shock-wave (SW) experiments have been 
carried out for Nb up to 124~GPa, with the bcc phase being the only phase 
observed along the Hugoniot \cite{21}. Here, in order to advance our knowledge 
of the phase diagrams of transition metals, we report ab initio calculations 
on the Nb phase diagram to 450~GPa and X-ray diffraction (XRD) experiments 
on Nb to 120~GPa and 6000~K. 

\section*{RESULTS}


\section*{Experimental}

XRD measurements reported here are from seven runs carried out on fresh 
samples to avoid the influence of possible chemical reactions on the results. 
They were performed at beamline ID27 of the European Synchrotron Radiation 
Facility (ESRF). See Methods section for both experimental and computational 
details. 

Fig. 1 shows a selection of XRD patterns collected at 38 and 117 GPa at 
different $T$s. They were obtained after integration of the charge-coupled 
device (CCD) images as a function of $2\theta .$ At 38~GPa NaCl was 
the pressure medium and at 117~GPa it was MgO. A normalized reference 
background has been subtracted from them. In order to do so, we used 
a measurement of the scattering from the DAC probing at a point of 
the pressure chamber where the sample was not present \cite{34}. At 38~GPa, 
as $T$ increases, the peaks slightly move to lower angles due to thermal 
expansion. We do not observe any other change in the XRD patterns up to 
2800~K. In particular, only the bcc phase of Nb and the B2 phase of NaCl are 
present. Bragg peaks of NaCl are detected even when the $T$ of the sample 
exceeds the melting $T$ $(T_{\rm m})$ of NaCl \cite{35}. This occurs because 
the NaCl in contact with the diamond anvils is at a lower $T$ than the sample, 
due to the high thermal conductivity of diamond. Thus, when the surface $T$ 
of Nb reaches $T_{\rm m}$ of NaCl, the latter is probably only partially 
molten, in the region where it is in contact with Nb. 
At 2800~K a typical powder diffraction pattern starts changing into 
a single-crystal-like one. Above that $T$, we observe rapid changes in 
the relative intensities of the Bragg peaks in consecutive diffraction 
patterns which are due to the recrystallization of the sample. This 
phenomenon has also been observed in other metals under HP conditions 
at several hundred degrees below $T_{\rm m}$ \cite{26,12,Mo,Pt,V}. 
It is probably caused by an increase in the grain size in polycrystalline 
material at HT. After this microstructural change, permanent reorientation 
of the crystal occurs, which causes changes in the relative intensities 
of the Bragg peaks (see Fig.~1(a)). This phenomenon is usually 
described in the literature as fast recrystallization (FR), and we will 
use this terminology in what follows. We note that in the case of Nb, 
an alternative (qualitative) description of this phenomenon can be 
given in terms of a transition to a virtual solid phase which becomes 
energetically competitive with bcc-Nb at the experimental $P$-$T$ 
conditions; see Results section for more detail. As a consequence, 
single-crystal spots appear and disappear randomly in subsequent XRD 
exposures. We observe similar behavior at 117~GPa. In this case $T_{\rm m}$ 
of MgO is higher than the maximum $T$ in the experiment \cite{36}, and 
therefore MgO remains in the solid B1 phase. In this experiment, the FR of 
Nb is observed at $T$ higher than 3400~K. The recrystallization described 
above has been observed at the seven $P$s at which we carried out experiments 
at $T$ several hundreds of degrees lower than the $T$ where the Bragg peaks 
of Nb disappear (probably due to melting, as discussed below). 

At 38 GPa, upon further heating to 3450 K we observe the disappearance of 
the Bragg peaks of Nb and the appearance of diffuse rings. At 117~GPa the same 
effect is detected at 4900~K. Such phenomena have previously been attributed 
to a signature of the onset of melting \cite{37}. However, this observation 
alone could lead to an overestimation of $T_{\rm m}$ \cite{Mo,Pt}. Likewise, 
the $T$ plateau, another signature of melting, may correspond to $T$ 
higher than the actual $T_{\rm m},$ which is the case for titanium, as 
discussed below. Therefore, the experimental $T_{\rm m}$s reported here, 
although definitely in the neighborhood of the actual melting curve 
of Nb, should only be considered approximate $T_{\rm m}$s. 

The diffuse rings at 3450~K are seen in Fig.\ 1 as two broad bands centered 
at $2\theta =10^\circ $ and $14^\circ .$ Qualitatively similar behavior has 
been observed in all the experimental runs. In the experiment carried out at 
117~GPa, the emergence of the diffuse scattering and the disappearance of 
bcc-Nb Bragg peaks is detected at 4900~K. In this case, the bump centered at 
$2\theta =10^\circ $ is broader than that at 38~GPa and more intense than 
the second bump. This qualitative difference in the diffuse scattering can be 
due to the fact that at 38~GPa NaCl may be also partially molten, as described 
above. Following the literature \cite{12,13,16,26,27}, we assume that 
the $T$ at which the phenomenon described above occurs corresponds to 
$T_{\rm m}$ of Nb. However, it should be noted that the detection of melting 
by XRD at high $P$ is not straightforward because the intensity of the diffuse 
scattering from the liquid is relatively low. In addition, the rapid change 
in the laser absorption by the sample during melting hinders the determination 
of $T_{\rm m}.$ Another cause of uncertainty may be an imperfect alignment of 
the x-ray beam and the hotspot during the whole heating cycle, although we 
tried to maintain this alignment to the best of our ability. 

Another important issue in the experiments is the possible occurrence of 
chemical reactions between the sample and the pressure medium or the sample 
with carbon that diffuses out of the diamond anvils. Usually, chemical 
reactions lead to a runaway heating \cite{38} which was not observed in our 
experiments. In addition, after each heating cycle, samples were optically 
inspected to check that there were no obvious signs of chemical reactions or 
oxidation on Nb surfaces. The only changes we visually detect on the sample 
surface after heating are textural changes and, in four of the seven runs, 
the formation of holes. A reasonable explanation is that the textures are 
a consequence of solidification upon quenching $T$, and that the holes are 
formed during melting \cite{39}. On the other hand, chemical reactions usually 
lead to the appearance of additional Bragg peaks (e.g., the formation of 
tantalum carbide in DAC experiments on Ta \cite{11}) that can be detected in 
the XRD patterns measured after reducing $T$ to room temperature (RT), which 
is not the case within the signal-noise ratio in our experiments. This and 
the rest of the observations summarized above suggest that no chemical 
reaction took place during the experiments. In any event, even if a chemical 
reaction occurred, its products would be undetectable by the current XRD 
methods and thus should not influence the reported results; a similar 
statement is also made in a paper on Ta \cite{11}. 
In two recovered samples we performed energy-dispersive X-ray analysis on 
a scanning electron microscope (Phillips XL-30) which allows the detection 
of light elements, and we did not find any detectable traces of either O or 
C (they can be detected in quantities as low as 1.0 wt\%). Other techniques 
sensitive to the presence of C, such as Raman spectroscopy, were not used to 
check for C contamination. However, as discussed above, we do have enough 
evidence to exclude the formation of a sizable amount of either Nb oxides or 
carbides. This conclusion is consistent with the fact that in XRD patterns we 
cannot identify Bragg peaks coming from any of the known Nb oxides or carbides 
\cite{40,41}. 


Solid-solid and solid-liquid transitions can be identified by the presence 
of a plateau in the sample $T$ when it is measured as a function of 
the increased laser heating power \cite{11,16}. However, it has been recently 
proposed that other phenomena can also cause $T$ plateaus \cite{plateaus}. 
In our case, we observe such features at two $T$s. One, when the sample $T$ 
reaches the onset of the recrystallization process described above, and 
the other, when the Nb Bragg peaks disappear. Fig.\ 2 shows the temporal 
evolution of the sample $T$ measured at 50~GPa. Note that the applied laser 
power increases linearly with time in our working regime. Two marked plateaus 
can be clearly identified in the figure, one at 2900~K, which starts 
simultaneously with the onset of recrystallization, and the other at 3650~K, 
which occurs after the disappearance of the Nb Bragg peaks that we associate 
with melting. Then, just like for both V \cite{V} and MgO \cite{MgO}, the $T$ 
plateaus observed in our experiments on Nb can be correlated with phase 
transitions. 

It is important to note here that at the five different $P$s in our 
experiments up to 50~GPa at $T$ below $T_{\rm m}$ but $\sim 200$-300~K above 
the recrystallization $T$ we observe the appearance of extra XRD peaks. 
After their first appearance, the new peaks are randomly observed as $T$ 
increases to $T_{\rm m}.$ These new peaks are seen in some XRD patterns; 
other XRD patterns only exhibit the peaks that are associated with bcc-Nb. 
For instance, Fig.~1 does not show any of the Pnma-Nb peaks above 
the corresponding bcc-Pnma transition T. This minimizes the possibility of 
a chemical reaction (which has already been ruled out based on the above 
considerations), which would otherwise lead to the appearance of new 
permanent peaks. A notable fact 
is that the extra Bragg peaks are found in the heating cycle only after 
$T$s higher than those of the first $T$ plateau are reached. 
Fig.~3 shows XRD patterns which exhibit this phenomenon. In both patterns, 
measured at (38~GPa, 3100~K) and (50~GPa, 3300~K), the extra peaks can be 
clearly identified. A possible explanation for this result is the existence 
of a stable or metastable HP-HT phase of Nb, similar to those proposed 
recently for both Mo \cite{8} and Ta \cite{2,4,31}. We made an attempt 
to index the XRD patterns where extra peaks are present, but we could not 
succeed assuming an extra solid phase of Nb, in addition to B2-NaCl, to be 
one of those that have been mentioned in the literature in connection with 
transition metals (from the previous section). However, ab initio 
quantum molecular dynamics (QMD) simulations helped us understand these 
XRD patterns. As we discuss in more detail below, in Nb at HP-HT 
ab initio calculations predict the existence of an orthorhombic 
structure that becomes thermodynamically more stable than bcc with increasing 
$T.$ This orthorhombic structure found in Nb is similar to that found in Ta 
\cite{2,31}. It belongs to space group Pnma and has four atoms per cell 
(bcc has two). In this structure Nb atoms are located at Wyckoff's 4c 
sites, $(x, 1/4, z),$ $(1/2-x, 3/4, 1/2+z),$ $(-x, 3/4, -z),$ and 
$(1/2+x, 1/4, 1/2-z),$ where $x\approx 0.13$ and $z\approx 0.35$ (and 
$b/a\sim 0.9$ and $c/a~\sim 0.5;$ see Fig.\ 4), according to calculations. 
By considering the structure determined theoretically, we were able to index 
all the extra peaks. According to the indexation (we assumed that $x=0.1325$ 
and $z=0.3444$ in either case), the unit-cell parameters of the Pnma 
structure at 38~GPa and 3100~K are $a=5.118(7)$~{\AA }, $b=4.526(7)$~{\AA }, 
and $c=2.712(4)$~{\AA }. At 50~GPa and 3300~K, the corresponding parameters 
are $a=5.021(9)$~{\AA }, $b=4.513(8)$~{\AA }, and $c=2.681(5)$~{\AA }. 
Additionally, we have been able to make decent Rietveld refinements of 
the XRD patterns shown in Fig.\ 3 by assuming the coexistence of the bcc 
and orthorhombic phases of Nb and considering also the presence of B2-NaCl. 
The residuals of the refinement are shown in Fig.\ 3. The small residuals 
of the refinement suggest that the coexistence of two HP-HT phases in Nb is 
a plausible explanation for our observations. The $R$-factors obtained from 
the refinement of the XRD pattern measured at 38~GPa and 3100~K are 
$R_{\rm p}=3.68$\% and $R_{{\rm wp}}=5.91$\%. Similar reliability of the fit 
parameters is achieved at 50~GPa and 3300~K. We note here that we found 
the same evidence for the possible existence of an orthorhombic HP-HT phase 
in Nb in the five experiments performed up to 50~GPa independent of 
the pressure medium used in these five experiments (NaCl or KBr; other 
pressure media Al$_2$O$_3$ and MgO were used at higher $P$) which also reduces 
the possibility that the extra peaks found in these experiments are due to 
a chemical reaction between the Nb sample and the pressure medium. 
Unfortunately, the rapid recrystallization of the sample makes it impossible 
to detect a pure orthorhombic phase in our experiments. As a consequence, we 
could not determine the exact locus of the corresponding solid-solid phase 
boundary. We note that when the sample is cooled down to RT, its quenched 
XRD patterns do not exhibit any peaks or any amorphous bands that can be 
associated with the orthorhombic structure. Therefore, the proposed 
bcc-Pnma solid-solid phase transition is reversible. 

The unit-cell volume determined for the orthorhombic structure at 38 GPa 
(15.705~{\AA }$^3$ atom$^{-1}$) is $\sim 1$\% smaller than that of bcc-Nb at 
the same $P$ (15.852~{\AA }$^3$ atom$^{-1}$). This small volume change at 
the bcc-Pnma transition suggests that the corresponding solid-solid phase 
boundary is rather flat, in view of the Clausius-Clapeyron (CC) formula 
$dT/dP=\Delta V/\Delta S,$ and should be similar to the flat melting curves 
in the former DAC experiments \cite{14,13,16,39}. Indeed, as seen in Figs.\ 
5,6, the bcc-Pnma transition boundary is almost flat up to $\sim 50$~GPa 
and starts increasing slowly at higher $P.$ At 50~GPa, the unit-cell volume 
determined for Pnma-Nb (15.188~{\AA }$^3$ atom$^{-1}$) is almost identical to 
that of bcc-Nb [15.186~{\AA }$^3$ atom$^{-1}$; $a_{\rm bcc}=3.120(3)$~{\AA }]. 

The calculated positions of the peaks assigned to Pnma-Nb and bcc-Nb are 
indicated by the corresponding vertical ticks in Fig.\ 3 (where the positions 
of Bragg NaCl peaks are also shown). Note that one of them overlaps with 
the most intense peak of bcc-Nb, and another one with the most intense peak 
of NaCl. On the other hand, we have not observed any diffraction pattern with 
the Pnma-Nb peaks alone. There were always peaks of the bcc phase overlapping 
with those of the orthorhombic phase, which indicates that the two phases 
effectively coexist. Such coexistence can be caused by either a FR or radial 
$T$ gradients, or both. Here a FR means a continuous transformation 
of one solid phase to another, which is described in more detail below. 
Radial $T$ gradients are always present in a laser-heated DAC; in 
our experiments they were minimized but could not be completely eliminated, 
hence the surface of the sample can be $\sim 200$-300~K hotter than its bulk, 
in which case XRD picks up a combined signal from the two coexisting solid 
phases belonging to different sections of the sample that have different $T$s. 
In Fig.\ 5 we summarize the results obtained in the seven experimental runs. 
It can be seen that the $T$ difference between the onset of recrystallization 
and melting increases with $P,$ which could be interpreted as the scaling 
of the $T$ of recrystallization with the $T$ of melting. We also indicate 
the five $P$-$T$ points (black stars) where the orthorhombic Pnma structure 
of Nb is detected. 

\section*{Theoretical}


\subsection*{Ab initio melting curve of Nb}

For the calculation of the melting curve of bcc-Nb a 432-atom 
($6\times 6\times 6$) supercell with a single $\Gamma $-point is used. Full 
energy convergence (to $\stackrel{<}{\sim }1$~meV atom$^{-1}$) is checked 
for each simulation. The six bcc-Nb melting points are listed in Table~1. 
For each of them, ten NVE (fixed total number of atoms N, system volume V, and 
total energy E; V corresponds to one of the six densities from Table~1) runs 
of 10000-15000 time steps of 1.5~fs each, with an increment of the initial $T$ 
of 62.5~K are carried out. Such a small $T$ increment is chosen to ensure 
the high accuracy of the results, since the error in $T_{\rm m}$ is half of 
the increment of the initial $T$ \cite{Os}. Thus, the errors of our six values 
of $T_{\rm m}$ are $\sim 30$~K, which is 1.25\% for the first point and less 
than  1\% for the remaining five. The $P$ errors are also negligibly small: 
$\stackrel{<}{\sim }0.5$~GPa for the first point, and 1-2~GPa for 
the remaining five. Hence, our melting results on bcc-Nb are highly accurate 
indeed. 

Fitting the Simon-Glatzel (SG) form \cite{42} to the six bcc-Nb points 
gives the melting curve of bcc-Nb ($T_{\rm m}$ in K, $P$ in GPa), 
\beq
T_{\rm m}(P)=2750\,\left( 1+\frac{P}{22.6}\right) ^{0.30}\!\!,
\eeq
the initial slope of which, $dT_{\rm m}(P)/dP=36.5$~K/GPa at $P=0,$ is 
in excellent agreement with 36~K/GPa from isobaric expansion measurements 
\cite{20}. Another theoretical value for this slope, 37.7~K/GPa 
\cite{43-1,43-2}, is only slightly higher. Note that the initial 
$dT_{\rm m}/dP$ from classical MD simulations is $\sim 50$~K/GPa 
using the embedded-atom method \cite{44}, or $\sim 54$~K/GPa using 
the Mie-Lennard-Jones pairwise interatomic potential \cite{Nb-Tm}. 

According to our cold free energy (i.e., enthalpy) calculations, 
the structures that are closest to bcc energetically are A15 ($\beta $-W) 
and three orthorhombic structures described with space groups 
Pnma ({N\textsuperscript{\underline{o}}} 62), 
Cmcm ({N\textsuperscript{\underline{o}}} 63), and 
Cmca ({N\textsuperscript{\underline{o}}} 64). At low $P,$ all four are less 
stable than bcc-Nb by $\sim 0.2$~eV atom$^{-1}$, although A15 is nominally 
the closest one to bcc. For comparison, the hexagonal close-packed (hcp)-bcc 
and face-centered cubic (fcc)-bcc values are $\sim 0.3$ and 0.35~eV 
atom$^{-1}$, respectively. For each solid structure, the difference of its 
enthalpy and that of bcc increases with $P.$ For example, for Pnma-Nb 
(in~eV~atom$^{-1}$) $\Delta H=0.2241+4.602\cdot 10^{-4}P+3.016\cdot 10^{
-7}P^2.$ The above four structures closest to bcc were among those for which 
we simulated melting curves. It turns out that there is only one solid 
structure the melting curve of which crosses that of bcc-Nb at $P$ below 
100~GPa, which is therefore relevant to our experimental work. This structure 
is Pnma. The melting curves of the other two orthorhombic structures may 
become relevant at $P$ above 100~GPa, although additional more careful study 
may be required to clarify this point. As a matter of fact, the melting curve 
of bcc-Nb is the highest among those of all the solid structures that we 
considered, including A15. 

The melting curve of Pnma-Nb is calculated exactly the same way as that of 
bcc-Nb, namely, 10 runs per point, 10000-15000 time steps of 1.5~fs each per 
run, and an increment of the initial $T$ of 62.5~K. In this case, a 448-atom 
($4\times 4\times 7$) supercell with a single $\Gamma $-point is used, and 
again, full energy convergence (to $\stackrel{<}{\sim }1$~meV atom$^{-1}$) is 
checked for each simulation. The six melting points of Pnma-Nb are listed in 
Table~2. 

The melting curve of Pnma-Nb is described by the SG form (again, 
$T_{\rm m}$ in K and $P$ in GPa) 
\beq
T_{\rm m}(P)=2710\,\left( 1+\frac{P}{21.0}\right) ^{0.34}\!\!.
\eeq
It crosses the melting curve of bcc-Nb, Eq.\ (1), at the predicted 
bcc-Pnma-liquid triple point (5.9~GPa, 2949~K). 


Supplementary Figures 1,2, and 3,4 demonstrate the time evolution of $T$ and 
$P$ in the $Z$-method runs of the bcc-Nb melting point $(P,T)=(124,4820)$ and 
the Pnma-Nb melting point $(125,5240),$ respectively. These two points are 
chosen as examples and are shown as green and red diamonds in Fig.~5. 

\subsection*{bcc-Pnma solid-solid phase transition boundary}

We have also obtained the bcc-Pnma solid-solid phase transition boundary from 
the calculation of the total free energies of both structures using TDEP at 
four different $P$s. 

In Fig.\ 7 we show the phonon spectra of Pnma-Nb at six different $T$s (from 
very low $T$ to slightly above $T_{\rm m},$ at a density of 9.95~g~cm$^{-3}$). 
It is seen that all the structures are mechanically stable (no imaginary 
phonon branches), and the $T$-dependence of their phonon spectra is very weak. 
We found that Pnma-Nb is mechanically stable at all $P.$ Hence, its 
total free energy can be calculated and compared to that of bcc-Nb. Fig.\ 8 
shows the Pnma-bcc total free energy difference, $\Delta F,$ as a function of 
$T$ at four different $P$s. The four starting points lie on the curve that 
describes the $T=0$ Pnma-bcc enthalpy difference from the previous subsection. 
At each $P,$ it is fitted with a cubic polynomial having a positive root that 
defines the corresponding bcc-Pnma transition point. The four transition 
points, $(P,T)=(155,\,3623),$ $(215,\,3942),$ $(295,\,4448),$ and 
$(395,\,5141),$ define the Pnma-bcc solid-solid phase transition boundary. 
In what follows, we take the triple point to be $(6,2950),$ and the bcc-Pnma 
phase boundary is 
\beq
T(P)=2950-13.9\,(P-6)+10.9\,(P-6)^{1.1},
\eeq
which interpolates between the low-$P$ experimental data and high-$P$ 
theoretical data better than the quadratic fit. The four theoretical 
bcc-Pnma transition points and the phase boundary are shown in Fig.\ 6. 

\subsection*{Nb Principal Hugoniot}

In order to get another confirmation of the existence and the location of 
the bcc-Pnma solid-solid phase transition boundary derived theoretically 
in this work, we now turn to the experimental results on the principal 
Hugoniot of Nb. 

If the bcc-Pnma phase boundary shown in Fig.\ 6 is correct then the Hugoniot 
crosses it. The Nb principal Hugoniot that we calculated theoretically, 
$T(P)=293+0.22\,P^{1.93}\!,$ crosses the bcc-Pnma transition boundary given by 
Eq.\ (3) at 143.4~GPa. In this case, assuming that the bcc and Pnma portions 
of the Nb Hugoniot are described by straight segments $U_{\rm s}=a+b\,U_{
\rm p},$ the piecewise linear function $U_{\rm s}=U_{\rm s}(U_{\rm p})$ 
describing the experimental data has a breakpoint at $P\sim 145$~GPa. Fig.\ 9 
shows $(U_{\rm s}-U_{\rm p})$ vs. $U_{\rm p}$ instead of $U_{\rm s}$ vs. $U_{
\rm p};$ the data come from the Russian Shock Wave Database \cite{RSWD}. We 
have chosen $(U_{\rm s}-U_{\rm p})$ instead of $U_{\rm s}$ to illustrate 
a change in compressibility, since on the Hugoniot $U_{\rm s}/(U_{\rm s}-U_{
\rm p})=\rho /\rho _0$ \cite{JD}.  Each of the straight segments are of 
the form $U_{\rm s}-U_{\rm p}=a+(b-1)\,U_{\rm p}$ where $a=\sqrt{B_0/\rho _0}$ 
and $b=(1+B_0')/4$; $B_0$ and $B_0'$ are the adiabatic bulk modulus and its 
first pressure derivative \cite{JD}, both of which come from the corresponding 
EOS. The value of the ambient density, $\rho _0,$ is within a few percent of 
that of $\rho $ at $(T=0,\,P=0)$ which we also denote as $\rho _0$ below. 
We have calculated the $T=0$ equations of state of both bcc-Nb and Pnma-Nb 
which are shown in Fig.\ 10. Their third-order Birch-Murnaghan forms are 
$$P(\rho )=\frac{3}{2}\,B_0\left( \eta ^{7/3}-\eta ^{5/3}\right) \left[ 1+
\frac{3}{4}(B_0'-4) \left( \eta ^{2/3}-1\right) \right] ,$$ where $\eta =
\rho /\rho _0$ and $(\rho _0$ in g~cm$^{-3}$ and $B_0$ in GPa) $${\rm bcc-Nb:}
\;\;\;\rho _0=8.5696,\;\;B_0=170.5,\;\;B_0'=3.85,$$ $${\rm Pnma-Nb:}\;\;\;
\rho _0=8.5717,\;\;B_0=137.2,\;\;B_0'=4.65.$$ In each of the two cases, this 
EOS is expected to be reliable to $\sim 500$~GPa. Note that our value of 
$\rho _0$ for bcc-Nb is identical to the Nb experimental ambient density. 
Our set of $(B_0,B_0')$ for bcc-Nb is consistent with similar sets that can 
be found in the literature: $(168.8,3.73)$ \cite{Kinslow}, $(169,4.08)$ 
\cite{KMF}, $(168(4),3.4(3))$ \cite{23}, $(174.9(3.2),3.97(9))$ \cite{Zou} 
(the average of the four sets is $(169.5,3.9)).$ 

We also note that the finite-$T$ counterparts of the above two EOSs can be 
written approximately as $P(\rho, T)=P(\rho )+\alpha \,T,$ where $\alpha _{
\rm bcc}=3.8$~$\cdot $~$10^{-3}$ and $\alpha _{\rm Pnma}=4.3$~$\cdot $~$10^{
-3}.$ For example, for the two experimental unit cell volumes of Pnma-Nb that 
correspond to 38 and 50~GPa mentioned above, 15.705~{\AA }$^3$ atom$^{-1}$ 
and 15.188~{\AA }$^3$ atom$^{-1}$ (the corresponding densities are 9.823 and 
10.16~g~cm$^{-3}$), the Pnma-Nb finite-$T$ EOS with the corresponding set 
of $(B_0,\,B_0',\,\rho _0,\,\alpha )$ gives $P(9.823,\,3100)=39$~GPa and 
$P(10.16,\,3300)=49$~GPa, in good agreement with 38 and 50, respectively. 

With the above values of $\rho _0,$ $B_0$ and $B_0',$ the two straight 
segments shown in Fig.\ 9 are $U_{\rm s}-U_{\rm p}=4.4605+0.2125\,U_{\rm p}$ 
(bcc) and $U_{\rm s}-U_{\rm p}=4.0008+0.4125\,U_{\rm p}$ (Pnma). Agreement 
between the experimental data and the two segments is excellent. The segments 
cross each other at $U_{\rm p}=2.2985\approx 2.3$~km s$^{-1}.$ According to 
the Nb data of \cite{RSWD}, along the Nb Hugoniot 
$P(U_{\rm p})=39.0\,U_{\rm p}+10.3\,U_{\rm p}^2;$ 
hence, at the transition point $P(U_{\rm p}=2.3~{\rm km\;s}^{-1})=144.2$~GPa, 
in excellent agreement with the value 143.4~GPa found above. Thus, 
the bcc-Pnma Hugoniot transition point is at $\sim 144$~GPa.  Melting 
on the Hugoniot occurs at $\sim 200$~GPa ($U_{\rm p}\sim 2.9$~km s$^{-1}$), 
where the Hugoniot crosses the Pnma-Nb melting curve, as seen in Fig.~6. 
Hence, at the two highest data points in Fig.\ 9 the Nb is a liquid. These 
points lie slightly below the straight segment most likely because the bulk 
modulus of a liquid is slightly smaller than that of a solid. 

We note that a solid-solid transition on the Nb Hugoniot was also detected in 
classical molecular dynamics simulations by Germann \cite{Germann}. He found 
a transition from bcc to another solid phase at $U_{\rm p}\sim 2.0$~km s$^{
-1},$ rather close to the 2.3~km s$^{-1}$ found here. However, he associated 
the high-$PT$ phase with hcp rather than Pnma. This was done based on 
the radial distribution function (RDF) of the simulated structure emerging 
above the transition $P$ after shocking in the [110] direction. This RDF 
is noisy and cannot be relied upon to identify the high-$PT$ phase. 

\section*{The phenomenon of fast recrystallization}

The mechanism of fast recrystallization observed in our experiments on Nb 
can be qualitatively characterized using the approach that we developed to 
describe a fcc-14H solid-solid phase transition experimentally observed in 
iridium \cite{Ir}. 

The nonhydrostaticity that is always present in a DAC \cite{Ir} results in 
an effective increase of the total free energy $(F)$ of the sample by 
the amount $\Delta F=2VG\sigma ^2/B(3B+4G),$ where $V$ is volume, $B$ and $G$ 
are the bulk and shear moduli, respectively, and $\sigma $ is 
the nonhydrostatic stress. The use of the $PT$-dependent $V,$ $G$ and $B$ 
\cite{Nb-thermo}, as well as $\sigma =0.2+0.025\,P$ from 
\cite{Nb-sigma1,Nb-sigma2}, in the above expression leads to 
(in eV atom$^{-1}$), up to the quadratic terms in $P$ and $T$) 
$$\Delta F=8.4\cdot 10^{-5}\,P+2.3\cdot 10^{-6}\,P^2+9.9\cdot 10^{-5}\,T
-3.0\cdot 10^{-8}\,T^2.$$   When the Pnma-bcc total free energy deficit is 
compensated by an increase in $F_{\rm bcc}$ of $\Delta F,$ a bcc-Pnma 
transition starts. The family of quasi-parallel curves in Fig.~8, and 
the expression for the bcc-Pnma enthalpy difference from above render 
the following formula for the Pnma-bcc $F$ difference 
$$0.224+4.6\cdot 10^{-4}\,P+3.0\cdot 10^{-7}\,P^2-3.9\cdot 10^{-5}\,T
-1.1\cdot 10^{-8}\,T^2.$$  Equating this to $\Delta F$ yields 
the bcc-Pnma transition $T$ as a function of $P$ under nonhydrostatic stress, 
$$T=3632\,-\,2.354\,\sqrt{252500\,-\,3572\,P\,+\,19\,P^2},$$ which is very 
accurately approximated by $T=2450\,+\,7\,P$ up to $\sim 60$~GPa, in 
excellent agreement with the dashed line of FR in Fig.~5. This formulation 
is valid up to $\sim 60$~GPa; at higher $P,$ $\sigma =\sigma (P)$ flattens 
out \cite{Nb-sigma1,Nb-sigma2}, so that $\sigma /P\ll 1$ and $\Delta F\ll 
F_{\rm bcc},$ which means that the line of fast resrystallization gets very 
close to the bcc-Pnma thermodynamic transition boundary (TTB), again in 
excellent agreement with experiment. 

The above arguments strongly suggest the following mechanism of fast 
recrystallization. Under nonhydrostatic stress in a DAC, bcc-Nb becomes 
thermodynamically unstable and converts to Pnma-Nb, which results in 
the release of nonhydrostatic stress. This happens below TTB at $P \stackrel{
<}{\sim }60$~GPa and very close to TTB at higher $P$; see Fig. 5. But since 
Pnma is thermodynamically less stable than bcc-Nb at these $P$-$T$ conditions, 
it reverts back to bcc, which again converts to Pnma, and so on. A similar 
mechanism can drive FR above TTB which shows up as the bcc-Pnma coexistence. 
Indeed, as $T$ increases and the bcc-Pnma transition point is approached and 
eventually reached, bcc-Nb becomes thermodynamically unstable and converts 
into Pnma-Nb. Then, under nonhydrostatic stress, the total free energy 
advantage of Pnma over bcc is reversed, by the same mechanism that reverses 
the total free energy advantage of bcc over Pnma below TTB discussed above, 
so that Pnma converts to bcc, which reverts back to Pnma, and so on. 

The difference between the two cases of FR considered above, namely, below 
and above TTB, is that bcc-Pnma coexistence is only observed in the latter 
case but not in the former. The lowest-$P$ Pnma data point in Fig.~5 (a star 
at $\sim 2800$~K) is the only Pnma data point that may seem to contradict 
the above statement. We note, however, that taking into account 
the corresponding $T$ error bar of $\sim 200$-250~K may place this data 
point above the bcc-Pnma-liquid triple point, and therefore above TTB. 
The remaining four Pnma data points are very clearly on or above TTB. 

There may be a number of reasons for this to happen. First, although 
the nonhydrostaticity-based model for FR considered above predicts the $P$-$T$ 
location of the FR line below TTB correctly, a texture transition in bcc-Nb 
may be mainly (or solely) responsible for this FR, so that the corresponding 
XRD patterns exhibit the bcc phase only. Second, the appearance of Pnma may be 
suppressed for kinetic reasons, and it may be observed above TTB, as  
mentioned above, only because the surface of the sample is $\sim 200$-300~K 
hotter than its bulk, so that XRD picks up a combined signal from the two 
coexisting solid phases belonging to different sections of the sample that 
have different $T$s. In any event, the duration of the FR Pnma cycle is 
expected to be larger for FR above TTB than that for FR below TTB. Indeed, 
this duration directly depends on the kinetic factor 
$\exp \{\frac{\Delta F_{\rm bcc-Pnma}}{k_{\rm B}T}\}$, and $\Delta F_{
\rm bcc-Pnma}$ changes its sign across TTB. For instance, with values of 
$\Delta F=0.1$~eV atom$^{-1}$ and $T=3500$~K (so that $k_{\rm B}T$ corresponds 
to $\approx 0.3$~eV atom$^{-1}$) that are typical of Nb, $\exp \{-0.1/0.3\}
\approx 0.7$ but $\exp \{0.1/0.3\}\approx 1.4,$ a factor of two. This may be 
crucial for the observation of Pnma above TTB. There may be other reasons, 
too, which we do not mention here. It would be interesting to test the idea 
that a texture transition observed earlier on in Mo \cite{Mo}, and perhaps in 
a number of other metals including Nb, may be triggered by the underlying 
solid-solid phase transition, either to a real solid phase, like Pnma-Nb, or 
a virtual one, like fcc-Mo or hcp-Mo, both of which have been discussed in 
the literature (e.g., \cite{8}). The nonhydrostaticity-based model of FR 
considered in this work strongly suggests that this idea is meaningful. 

\section*{DISCUSSION}

The seven experimental $T_{\rm m}$s can be described by the SG equation 
$T_{\rm m}(P)=2750\,(1+P/48)^{0.45},$ where $T_{\rm m}$ is in K and $P$ in 
GPa, which is plotted in Fig.\ 5 as a solid black line. For this fit, 
the initial slope, 25.8~K/GPa, is significantly lower than the value 
36~K/GPa obtained from isobaric expansion measurements \cite{20}. On the other 
hand, the initial slope of the theoretical melting curve of bcc-Nb, Eq.\ (1), 
is 36.5~K/GPa, which is in excellent agreement with \cite{20}. We consider 
this theoretical melting curve to be the correct one for bcc-Nb. Above 
the bcc-Pnma-liquid triple point $P>6$~GPa) niobium melts from Pnma, and its 
melting curve is described by Eq.\ (2). The fit to the seven experimental 
points lies below Eq.\ (2) because of the three relatively low data points 
at $\sim 40$-60~GPa. The difference between their values of $T_{\rm m}$ and 
those predicted by Eq.\ (2) is $\sim 250$~K, or $\sim 7$\%, taking into 
account the upper error bar (perhaps it is even smaller because of 
$\sim 30$~K uncertainty in the values of the theoretical melting points). 
The remaining four experimental data points are consistent with Eq.\ (2). 
Next we suggest a possible reason for the experimental $T_{\rm m}$s 
being slightly lower than those predicted by the theory. 

As we have already mentioned above, the detection of melting by XRD at high 
$P$ may be problematic, and may cause uncertainty in the experimental melting 
$T.$ It turns out that a $T$ plateau, too, may be another source 
of uncertainty in experimental $T_{\rm m}.$ Although a $T$ plateau 
is considered to be an unambiguous signature of melting, there may be 
experimental conditions at which the plateau may correspond to $T$ different 
from the true $T_{\rm m}.$ Stutzmann et al. \cite{Ti} found that in 
the case of titanium the $T$ of plateaus decreases with $P,$ such that it 
differs significantly (by up to $\pm 1000$~K) from the $T_{\rm m}$ estimated 
using XRD data. They attribute this difference in $T$ to a decrease in thermal 
insulation due to the thinning of the pressure chamber with increasing $P,$ 
and suggest that in the case of Ti the $T$ plateaus cannot be used 
as a systematic diagnostic for the sample melting. In our case of Nb, 
the plateaus do not seem to be as drastically different from the XRD melting 
data as for Ti, but a relatively small difference, $\sim 200$-250~K, between 
the $T$ of the plateaus and the true $T_{\rm m}$ may occur, for reasons 
similar to those responsible for large differences in the case of titanium. 
Note that the locations of the $T$ plateaus may be also influenced by 
the FR detected in our experiments below melting. 

The theoretical bcc-Pnma solid-solid phase boundary is shown in Fig.\ 6. 
We would like to emphasize that this phase boundary is consistent with 
the experimental data points where evidence for the existence of the proposed 
HP-HT Pnma phase is found. Note that at low $P$ the bcc-Pnma phase boundary 
is quasi-horizontal, which corresponds to a very small volume change across 
the bcc-Pnma transition, in view of the CC relation. This is very nicely 
confirmed by our experimental results, as discussed above. 

The theoretically predicted existence of a HP-HT orthorhombic phase not only 
provides an explanation for the appearance of the extra Bragg peaks observed 
in our experiments, but also for the observed FR. Unfortunately, it is so 
fast that the orthorhombic phase cannot be isolated in our XRD, in which 
images were collected every two seconds. Therefore, our results call for 
the development of ultrafast XRD techniques for HP studies. 

To summarize, we have determined experimentally and theoretically the melting 
curve of Nb and found very good agreement between experiments and theory. 
We also found evidence for the existence of Pnma-Nb, the orthorhombic HP-HT 
phase of Nb. The existence of this phase provides an explanation not only 
for the appearance of extra Bragg peaks in the XRD patterns at HP-HT 
conditions, but also for the sample recrystallization observed several 
hundred K below $T_{\rm m}$. The polymorphism of Nb makes it an excellent 
candidate for further studies that will advance our understanding of 
the $P$-$T$ phase diagrams of transition metals. 

We conclude by making the following four points that summarize our combined 
experimental and theoretical study on Nb at HP-HT. 

1. FR observed in HP-HT experiments on transition metals, including 
the present case of Nb, may correspond to either an isostructural texture 
change (Mo \cite{Mo}, V \cite{V}) or a true solid-solid phase transformation 
(Re \cite{Re}, Nb). In either scenario, the former DAC experiments must have 
misinterpreted the corresponding FR lines as the flat melting curves. 

2. The appearance of bcc-Nb above the bcc-Pnma transition point detected 
by XRD is probably due to some (small) nonhydrostaticity or radial thermal 
gradients being present in our experiments, as discussed in more detail above. 


3. Our findings provide clear evindence for the topological similarity of 
the phase diagrams of Nb and Ta, its group 5B partner in the periodic table. 
In both cases an orthorhombic phase transition occurs, and the high-PT 
physical solid structure is Pnma. However, the location of the bcc-Pnma-liquid 
triple point in Ta corresponds to $P$ an order of magnitude higher than that 
in Nb \cite{31}. 

4. Our results call for the development of ultrafast HP XRD techniques. 

\section*{METHODS}

\subsection*{Experimental details}

Disks of Nb (99.99\% purity, Aldrich), 15-25~$\mu $m in diameter and 5~$\mu $m 
thick, were loaded into Almax-Plate DACs (200-280~$\mu $m culets). The samples 
were placed in the center of a~60-90~$\mu $m diameter hole of a rhenium gasket 
pre-indented to a thickness of 30~$\mu $m. Al$_2$O$_3,$ KBr, MgO, and NaCl 
were used as pressure-transmitting media and to thermally insulate the Nb 
from the diamonds. $P$s were measured before and after heating using 
ruby fluorescence \cite{22}, and are in agreement with those obtained from 
the equations of state (EOS) of Nb \cite{23} and the pressure medium 
\cite{24}. The Nb samples were heated on both sides with two infrared YAG 
lasers which have a maximum power of 100~W each. The laser beams were 
defocused to create hotspots of 15-20~$\mu $m in diameter over the surfaces 
of the sample. The power and focusing of the lasers were adjusted to minimize 
the difference of $T$s on both sides of the sample to less than 100~K. 
Spectral radiometry was used to determine $T$s on the sample surfaces 
by collecting the thermal emission signal from circular areas of 2~$\mu $m 
diameter in the center of the hot spots \cite{25}. $T$ gradients in 
the center of the hot spots were within 10~K~$\mu $m$^{-1}$ ($\Delta T\sim 
100$~K in an area of 10~$\mu $m in diameter). The x-ray beam 
($\lambda =0.3738$~{\AA }) was focused on a $2\times 2$ $\mu $m$^2$ area. 
X-ray induced fluorescence of the diamond or the pressure medium was used 
to align the x-ray beam with the hot spot. Samples were compressed at RT up
to the selected $P,$ and then $T$ was raised at constant load by gradually
increasing the power of the lasers. After reaching the highest $T$ of each run,
$T$ was quenched by a rapid decrease of the laser power to zero in less than
3~seconds. In all the experiments, a pure bcc phase of Nb was recovered, thus
confirming RT XRD studies \cite{23} and SW measurements \cite{21}. No extra
Bragg peaks were detected in the quenched XRD patterns. During each run, XRD
and thermal emission spectra were recorded simultaneously every 2 seconds.
XRD patterns were collected using a MAR CCD detector. The same approach was
successfully used previously to study Fe \cite{26,27}, Ta \cite{11} and Ni
\cite{12}. 
It allows one to resolve the structure and to establish the onset of melting. 
In order to do so, the diffraction patterns measured with the CCD detector 
were integrated as a function of $2\theta $ in order to obtain one-dimensional 
XRD profiles. The indexing and refinement of the powder XRD patterns were 
performed using the DICVOL \cite{29} and PowderCell \cite{30} programs.

\subsection*{Computational details}

We have carried out an extensive study of the phase diagram of Nb using 
the Z method implemented with VASP (Vienna Ab initio Simulation Package) 
in combination with the temperature-dependent effective potential (TDEP) 
method. The Z method was used to simulate the Nb melting curve. The Z method 
is described in detail in \cite{31,Os,32}. TDEP \cite{Hellman1,Hellman2} takes 
into account anharmonic lattice vibrations. TDEP was used to calculate full 
free energies of the solid phases of Nb. The calculations were based on 
density-functional theory (DFT) with the projector-augmented-wave (PAW) 
\cite{PAW} implementation and the generalized gradient approximation (GGA) for 
exchange-correlation energy, in the form known as Perdew-Burke-Ernzerhof (PBE) 
\cite{PBE}. Since the simulations were performed at high-$PT$ conditions, we 
used accurate pseudopotentials where the semi-core 4s and 4p states were 
treated as valence states. Specifically, Nb was modeled with 13 valence 
electrons per atom (4s, 4p, 4d, and 5s orbitals). Cold $(T=0)$ total free 
energies (enthalpies) were calcualted using unit cells with very dense 
k-point meshes (e.g., $50\times 50\times 50$ for bcc-Nb) for high accuracy. 
In all the non-cubic cases we first relaxed the structure to determine its 
unit cell parameters at each volume. Fig.\ 4 shows the unit cell parameters, 
including two internal coordinates $x$ and $z$, for Pnma-Nb as functions of 
the lattice constant. Finite-$T$ total free energies were calculated (with 
TDEP) using supercells of order 400-500 atoms. Supercells were constructed 
based on the corresponding unit cells. Using the Z method, we calculated 
the melting curves of bcc-Nb as well as many other (virtual) solid structures 
of Nb which have been mentioned in the literature in connection with 
transition metals: all the close-packed structures with different layer 
stacking (fcc, hcp, dhcp, thcp, 9R), open structures (simple cubic, A15, 
hex-$\omega $), and different orthorhombic structures. Again, in all 
the non-cubic cases we relaxed the structure to determine its unit 
cell parameters; those unit cells were used for the construction of 
the corresponding supercells. We used systems of 400-500 atoms in each case. 
We did not consider the issue of mechanical stability of the other solid 
structures of Nb at low $T.$ We followed the logic that if the melting curve 
of a solid structure exists, this structure is certainly mechanically stable 
at a $T$ close to the corresponding melting point, $T_{\rm m},$ regardless of 
its mechanical stability at low $T.$ The melting curves were obtained for all 
the structures that we dealt with, except the simple cubic one, which would 
disorder during every molecular dynamics run and thus turned out to be 
mechanically unstable at all $T$. 

\section*{Data Availability}

All relevant data that support the findings of this study are available 
from the corresponding authors upon request. 


\section*{Acknowledgements}

The authors acknowledge the European Synchrotron Radiation Facility for 
the provision of synchrotron beamtime at the beamline ID27. Research 
financed by Spanish Ministerio de Ciencia, Innovaci\'{o}n y Universidades 
(Grants {N\textsuperscript{\underline{o}}} MAT2016-75586-C4-1-P, 
PGC2018-097520-A-100, and RED2018-102612-T), and Generalitat Valenciana 
(Prometeo/2018/123 EFIMAT). D.S.P. acknowledges the Spanish government 
for a Ramon y Cajal RyC-2014-15643 grant. 
This work was performed under the auspices of the U.S. Department of Energy 
by Lawrence Livermore National Laboratory under contract DE-AC52-07NA27344. 
QMD simulations were carried out on the Los Alamos IC clusters Conejo, 
Mapache, Pinto, and Badger. 

\section*{Author contributions}
D.E. designed the study; M.M. expedited the provision of the beamline; D.E, 
S.M., D.S., H.C., M.M. performed experiments; D.E., S.M., D.S., H.C. analyzed 
data; L.B., D.P. carried out QMD simulations; L.B., S.C., M.Mc., J.P. carried 
out theoretical calculations; S.S. carried out TDEP calculations; D.E., L.B. 
wrote the paper with contributions from all authors. 

\section*{Additional information}
{\bf Supplementary Information} accompanies this paper and can be found 
at https://www.nature.com/commsmat/ 

\subsection*{Competing financial interests:} The authors declare no competing 
financial or non-financial interests.


\begin{figure}
\begin{center}
\hspace*{-2.75cm}
\includegraphics[width=8.2in]{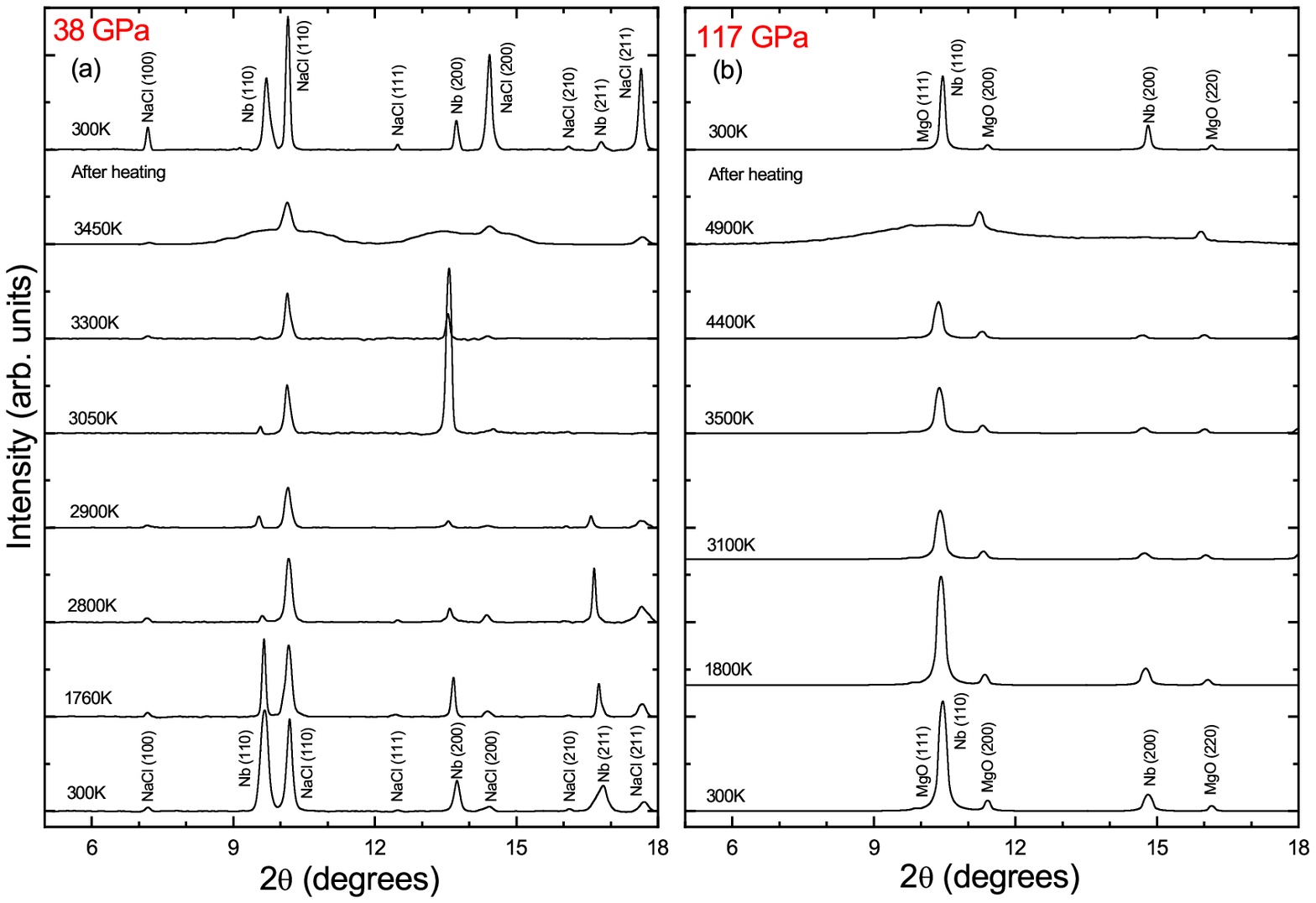}
\end{center}
\vspace*{-0.25cm}
\caption{\label{Figure 1}X-ray diffraction (XRD) patterns collected at 38 and 
117 GPa at different temperatures. Temperatures are indicated. Nb, NaCl, and 
MgO Bragg peaks are labeled. In order to highlight the detection of 
the recrystallization and melting phenomena at 38 GPa we have excluded XRD 
patterns where the orthorhombic Pnma structure is detected to coexist with 
body-centered cubic bcc-Nb; examples of such bcc-Pnma coexistence are shown 
in Fig.~3.}
\end{figure}

\begin{figure}
\begin{center}
\includegraphics[width=6.25in]{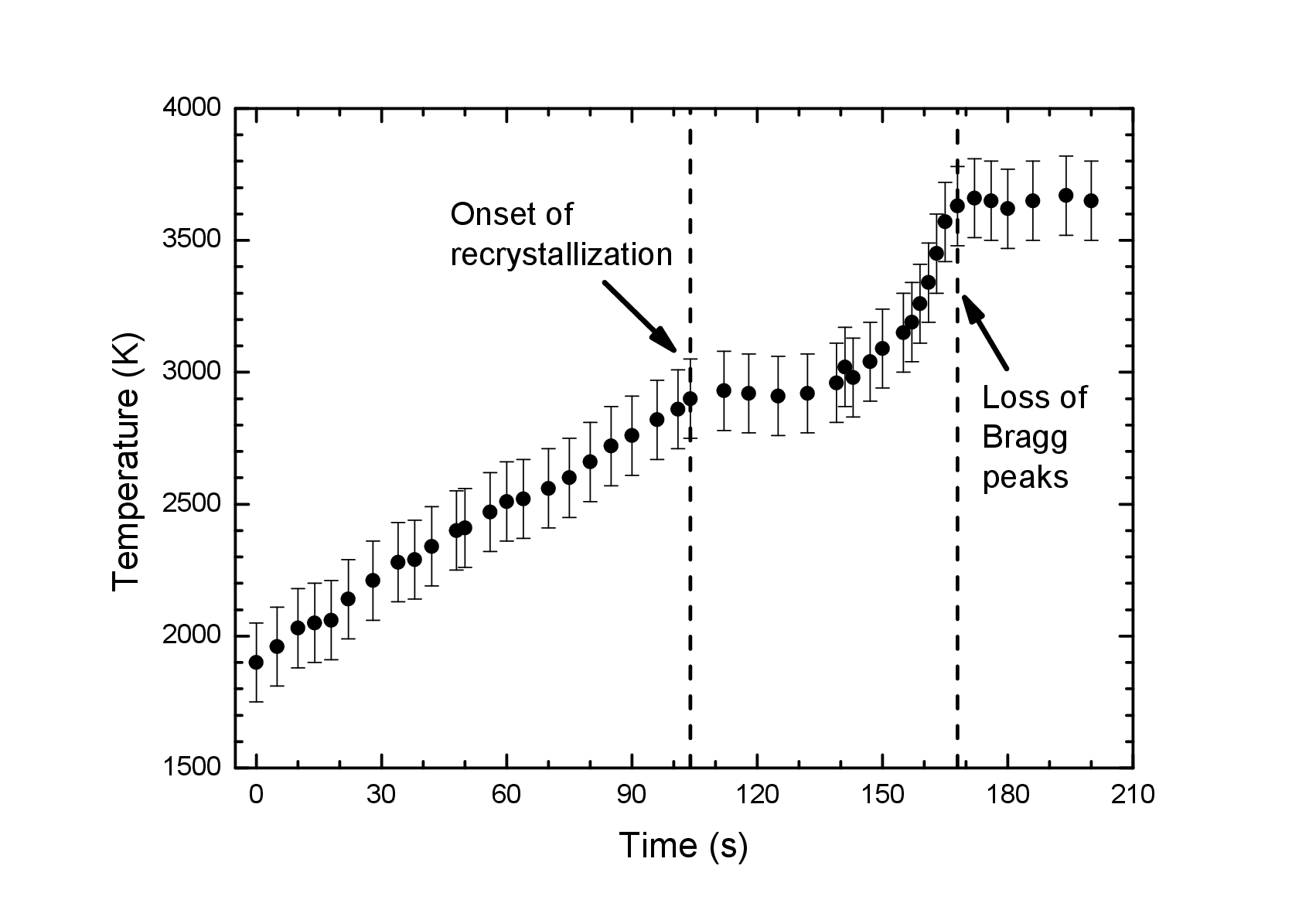}
\end{center}
\vspace*{-0.25cm}
\caption{\label{Figure 2}The Nb temperature measured as a function of time 
at 50~GPa. The vertical lines indicate the beginning of the two plateaus.}
\end{figure}

\begin{figure}
\begin{center}
\hspace*{-4.5cm}
\includegraphics[width=9.5in]{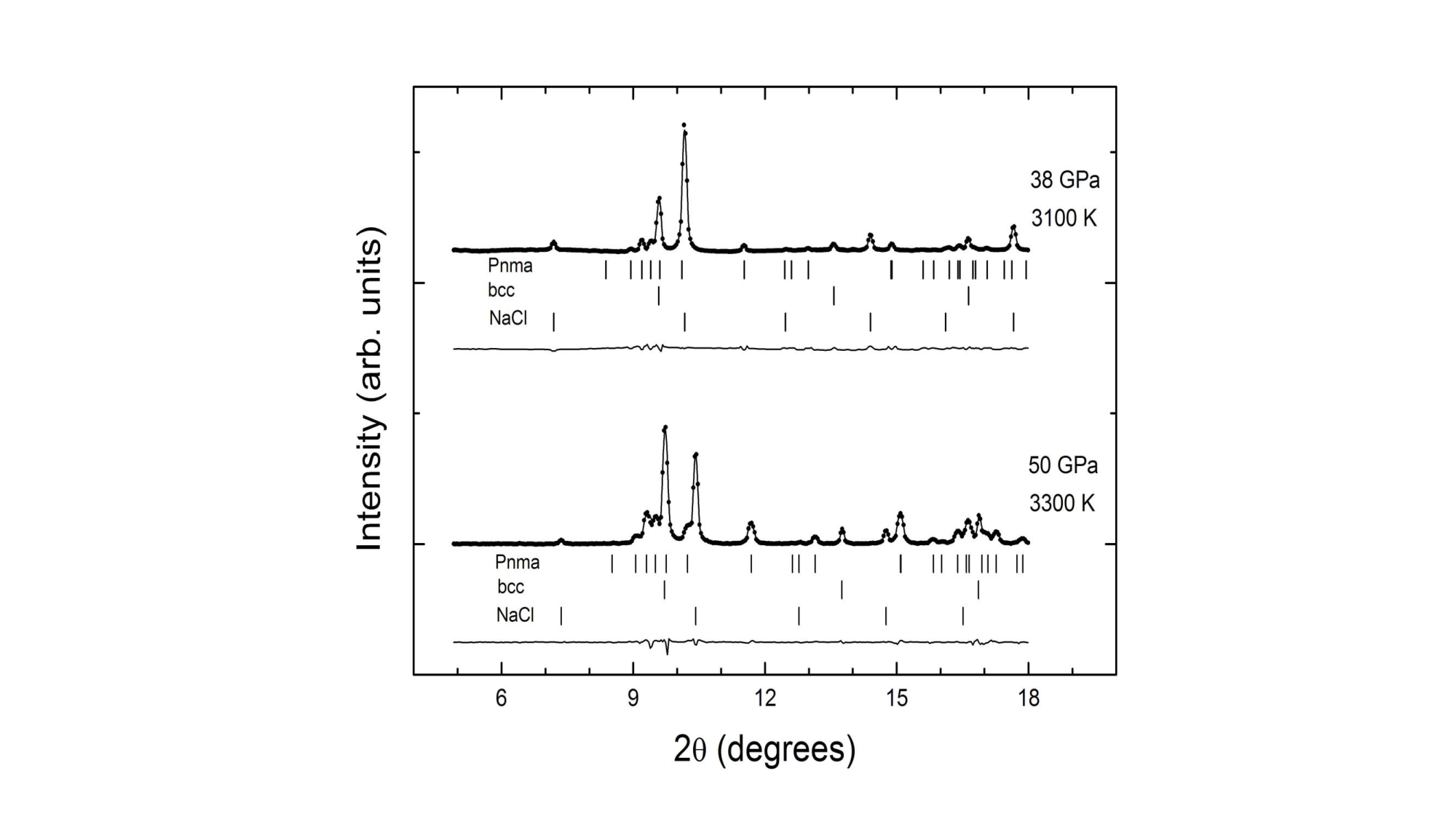}
\end{center}
\vspace*{-0.25cm}
\caption{\label{Figure 3}X-ray diffraction (XRD) patterns collected at 38 GPa 
and 3100 K, and at 50 GPa and 3300 K. The experimental XRD patterns are shown 
with dots and the Rietveld refinements and the residuals with solid lines. 
Ticks show the calculated positions for orthorhombic Pnma-Nb, body-centered 
cubic bcc-Nb, and B2-NaCl.}
\end{figure}

\begin{figure}
\vspace*{-0.5cm}
\begin{center}
\includegraphics[width=5.15in]{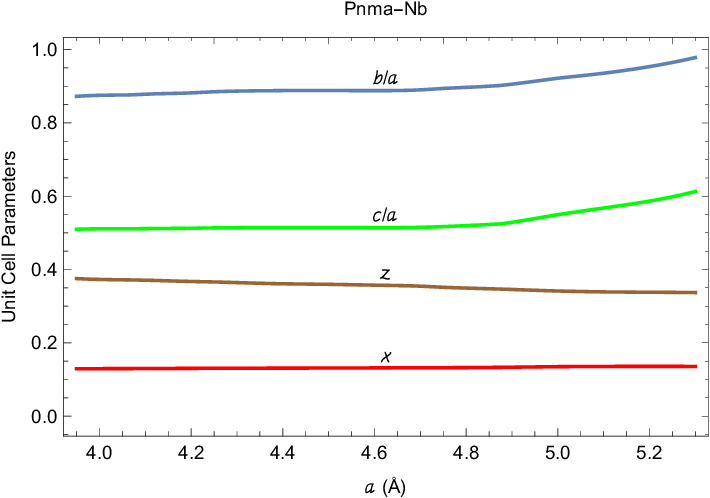}
\end{center}
\vspace*{-0.25cm}
\caption{\label{Figure 4}The unit cell parameters of orthorhombic Pnma-Nb, 
including two internal coordinates, $x$ and $z,$ as functions of the lattice 
constant $a,$ at zero temperature. The corresponding pressures span 
the $\sim (0,1000)$~GPa interval.}
\end{figure}

\begin{figure}
\begin{center}
\includegraphics[width=6.25in]{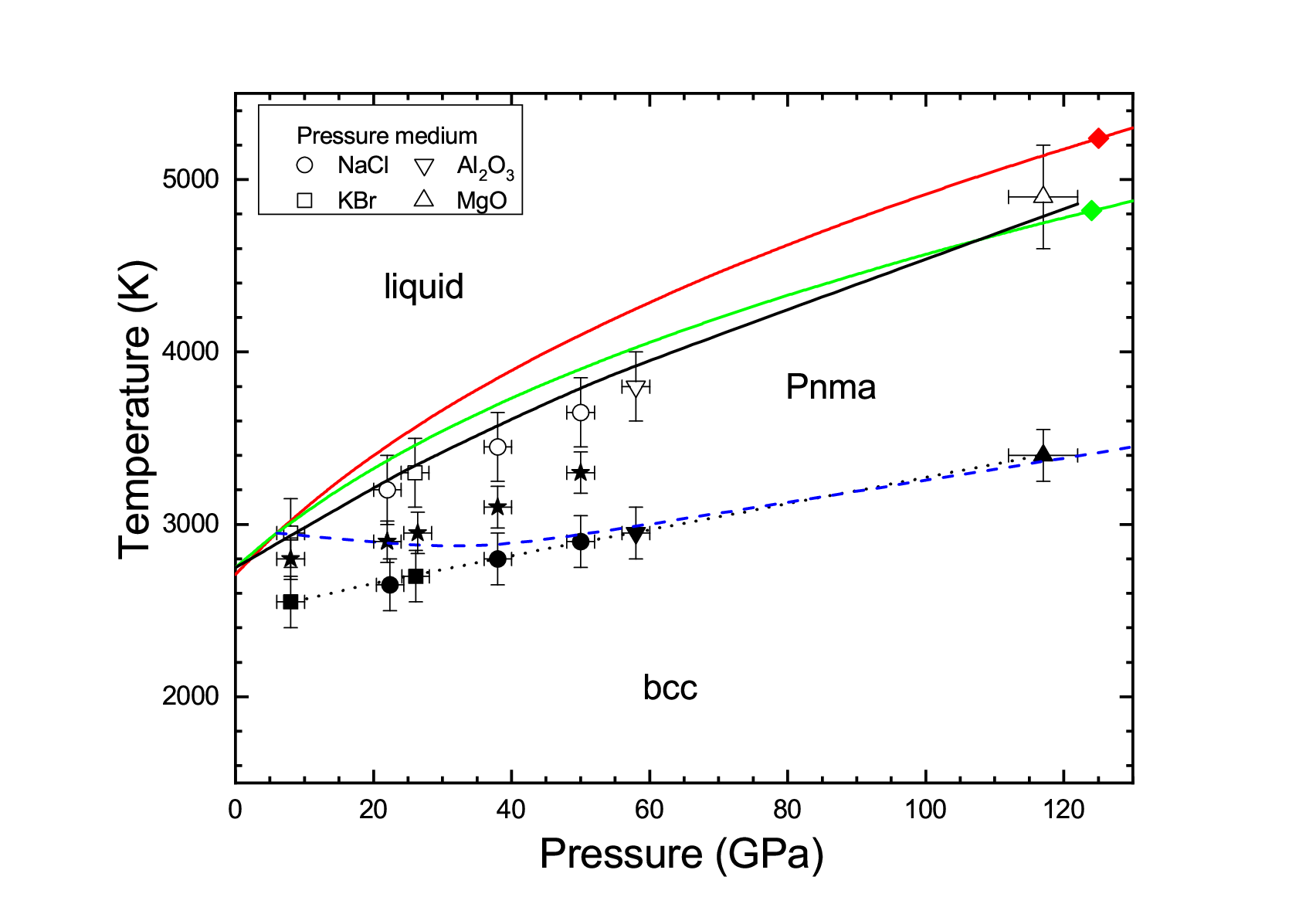}
\end{center}
\vspace*{-0.25cm}
\caption{\label{Figure 5}The low-pressure portion of Figure 6 including 
the experimental results of this work. Different symbols correspond 
to experiments done using different pressure media (see inset). Black solid 
symbols and black dotted line: the onset of recrystallization. Black empty 
symbols: the onset of melting. Black solid line: the Simon-Glatzel fit to 
the experimental melting points. Green and red lines: the quantum molecular 
dynamics (QMD) melting curves of body-centered bcc-Nb and orthorhombic 
Pnma-Nb, respectively. Black stars are the pressure-temperature points 
where the Pnma structure is detected in experiment. Blue dashed line: 
the theoretical bcc-Pnma phase boundary from this work. The error bars 
of the QMD melting points (green and red diamonds) are smaller than 
the size of the corresponding symbol.}
\end{figure}

\begin{figure}
\vspace*{-1.0cm}
\begin{center}
\includegraphics[width=5.15in]{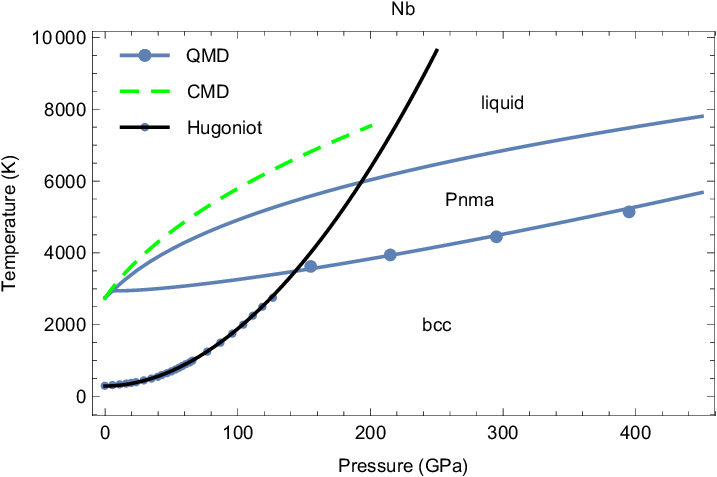}
\end{center}
\vspace*{-0.25cm}
\caption{\label{Figure 6}The ab initio phase diagram of Nb. The theoretical 
Hugoniot points of Ref.\ \cite{Nb-Hugoniot} are shown for comparison with our 
Hugoniot (black line). Also shown is the melting curve of Ref.\ \cite{Nb-Tm} 
obtained from classical molecular dynamics (CMD, green dashed line). The error 
bars of the quantum molecular dynamics (QMD) points are smaller than the size 
of the symbol.}
\end{figure}





\begin{figure}
\begin{center}
\includegraphics[width=6.25in]{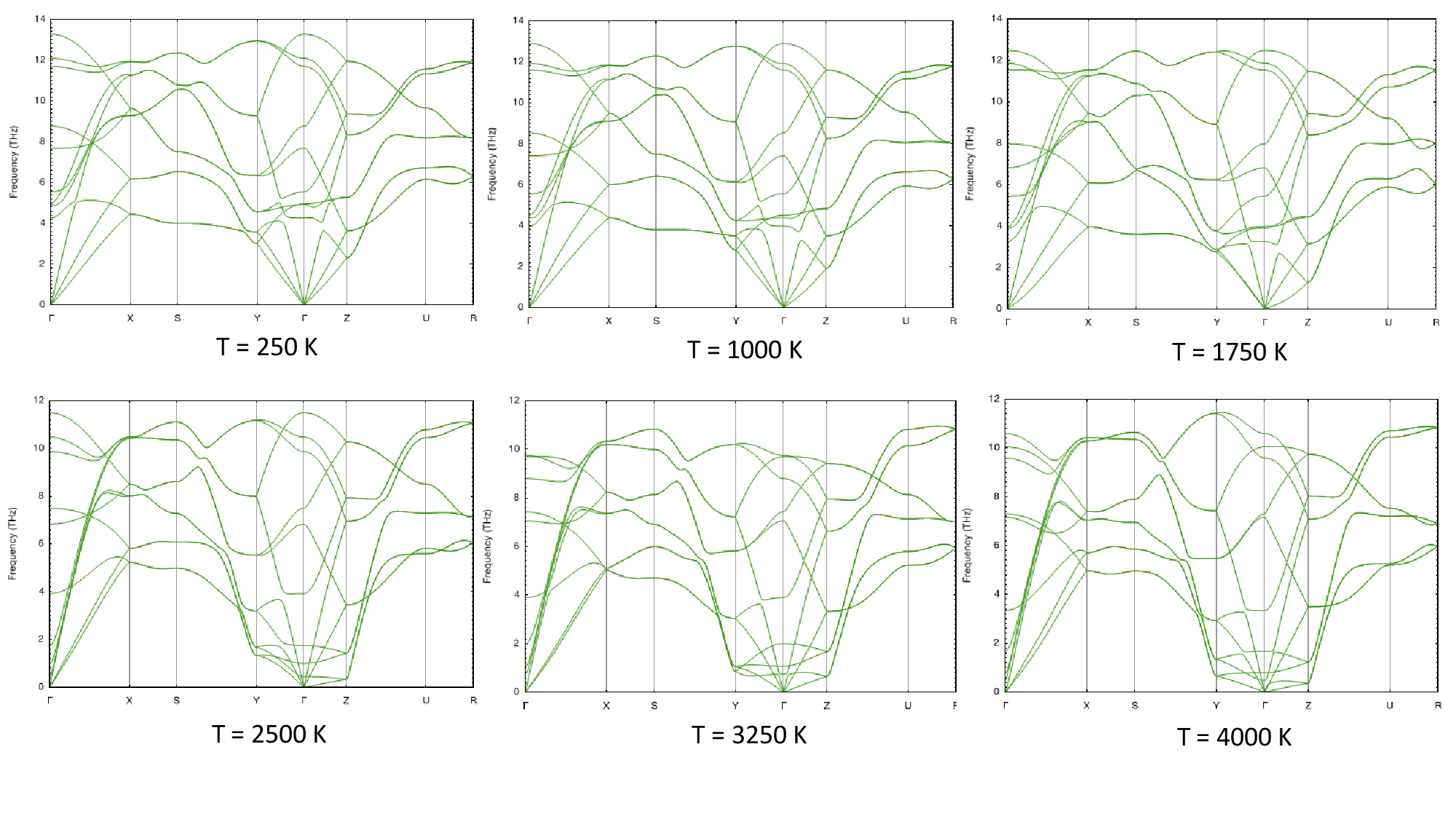}
\end{center}
\vspace*{-0.25cm}
\caption{\label{Figure 7}Phonon spectra of orthorhombic Pnma-Nb at six 
different temperatures at a fixed density of 9.95~g~cm$^{-3}.$}
\end{figure}

\begin{figure}
\vspace*{-1.0cm}
\begin{center}
\includegraphics[width=5.15in]{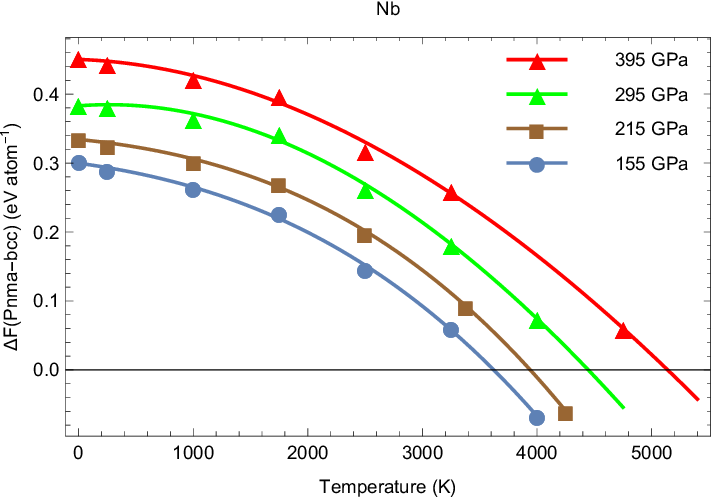}
\end{center}
\vspace*{-0.25cm}
\caption{\label{Figure 8}The difference between the total free energies 
of orthorhombic Pnma-Nb and body-centered cubic bcc-Nb as a function of 
temperature at four different pressures. Error bars are smaller 
than the size of the corresponding symbol.}
\end{figure}

\begin{figure}
\begin{center}
\includegraphics[width=5.15in]{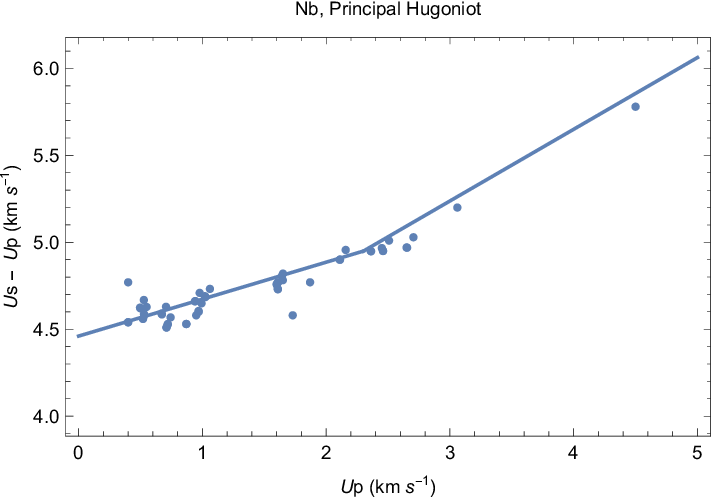}
\end{center}
\vspace*{-0.25cm}
\caption{\label{Figure 9}The difference of shock ($U_{\rm s})$ and particle 
($U_{\rm p}$) velocities as a function of $U_{\rm p}$ along the Nb Hugoniot. 
The two straight segments are the Hugoniots of body-centered cubic bcc-Nb 
(the lower one) and orthorhombic Pnma-Nb (the upper one) described by 
$U_{\rm s}-U_{\rm p}=a+(b-1)U_{\rm p}$ with the corresponding sets of $(a,b)$ 
parameters.}
\end{figure}

\begin{figure}
\vspace*{-0.5cm}
\begin{center}
\includegraphics[width=5.15in]{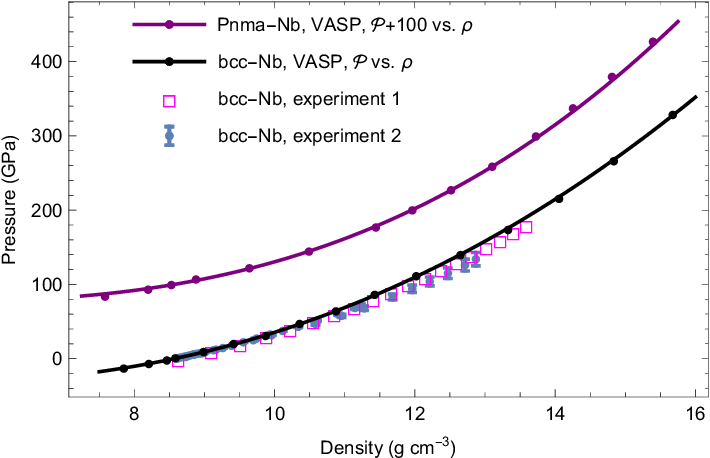}
\end{center}
\vspace*{-0.25cm}
\caption{\label{Figure 10}The zero temperature equations of state of 
body-centered cubic bcc-Nb and orthorhombic Pnma-Nb: VASP (black curve, bcc; 
magenta curve, Pnma) vs. the experimental results from Refs.\ \cite{Kinslow} 
(experiment~1, empty squares) and \cite{23} (experiment~2, bullets). 
The error bars of experiment~1 are smaller than the size of the symbol. 
The Pnma-Nb curve is shifted up by 100~GPa for clarity.}
\end{figure}



\begin{center}
\vspace*{0.5cm}
\begin{tabular}{|c|c|c|c|}
\hline 
lattice constant ({\AA }) & density (g cm$^{-3}$) & $P_{\rm m}$ (GPa) & 
$T_{\rm m}$ (K) 
 \\
\hline 
3.445 & 7.547 & -7.9 & 2400  \\
3.225 & 9.199 & 27.0 & 3460  \\
3.050 & 10.88 & 79.9 & 4330  \\
2.955 & 11.96 &  124 & 4820  \\
2.850 & 13.33 &  191 & 5390  \\
2.745 & 14.92 &  287 & 6050  \\
\hline 
\end{tabular}
\end{center}
\vspace*{0.25cm}
{\bf Table 1.} The six ab initio melting points of body-centered cubic bcc-Nb, 
$(P_{\rm m},\,T_{\rm m}),$ obtained from the~$Z$~method implemented with VASP. 
\vspace*{1.0cm}

\begin{center}
\vspace*{0.5cm}
\begin{tabular}{|c|c|c|c|}
\hline 
lattice constant ({\AA }) & density (g cm$^{-3}$) & $P_{\rm m}$ (GPa) & 
$T_{\rm m}$ (K)
 \\
\hline 
5.4270 & 8.143 &  6.5 & 2970  \\
5.1155 & 9.723 & 40.1 & 3890  \\
4.9290 & 10.87 & 76.4 & 4550  \\
4.7675 & 12.01 &  125 & 5240  \\
4.6390 & 13.04 &  180 & 5850  \\
4.5260 & 14.04 &  245 & 6430  \\
\hline 
\end{tabular}
\end{center}
\vspace*{0.25cm}
{\bf Table 2.} The six ab initio melting points of orthorhombic Pnma-Nb, 
$(P_{\rm m},\,T_{\rm m}),$ obtained from the~$Z$~method implemented with VASP.
\vspace*{1.0cm}

\end{document}